\documentclass[aps, prl, twocolumn,superscriptaddress]{revtex4-1}
\usepackage{amssymb,amsmath,bm,graphicx}
\usepackage{color}

\newcommand{\bsigma}{{\boldsymbol \sigma}}

\begin{document}

\title{Chiral Anomaly and Second Harmonic Generation in Weyl Semimetals}
  \author{A. A. Zyuzin}
\affiliation{Department of Theoretical Physics, The Royal Institute of Technology, Stockholm, SE-10691 Sweden}
 \affiliation{Department of Physics,
University of Basel, Klingelbergstrasse 82, CH-4056 Basel, Switzerland}
\affiliation{Ioffe Physico-Technical Institute, 194021 St. Petersburg, Russia}

\author{A. Yu. Zyuzin}
\affiliation{Ioffe Physico-Technical Institute, 194021 St. Petersburg, Russia}
\begin{abstract}
We study second harmonic generation in centrosymmetric Weyl semimetal with broken time reversal symmetry. We calculate electric current density at the double frequency of the propagating electromagnetic field in the presence of applied constant magnetic field, using the method of kinetic equation for electron distribution function. It is shown that the chiral anomaly contribution to second harmonic generation in the lowest order is linearly proportional to the applied magnetic field.
The limit when the chiral anomaly dominates over the Lorentz-type contribution to second harmonic generation is discussed. 
\end{abstract}


\maketitle

\textit{Introduction}. Dirac-Weyl semimetals are new theoretically predicted \cite{Weyl, Weyl1, Weyl2, Weyl3, Murakami} and experimentally discovered materials  
\cite{Xu613, PhysRevX.5.031013, CdAs, bib:WSM1, bib:WSM2, Kharzeev} 
extending the class of two-dimensional topological electronic states of matter to three dimensions \cite{Volovik, Balatsky_Dirac}. 

The electronic excitations in these systems behave as Dirac-Weyl fermions, which allows the study of relativistic quantum mechanics in the table-top condensed matter experiments. 
In particular the bulk band structure of Weyl semimetals is characterized by the presence of two or more nondegenerate linearly dispersing band-touching points, called Weyl points,
associated with chiral massless fermions in three-dimensional real space. These Weyl points behave as a sink or source for the Berry curvature in momentum space. 
The Weyl semimetal state can be realized by breaking time reversal or inversion symmetries in the system tuned 
to the vicinity of the topological insulator - normal insulator phase transition \cite{Anton_review}.

Weyl fermions give rise to the chiral anomaly \cite{PhysRev.177.2426, Bell}, which is the non-conservation of the number of particles of a given chirality induced by
the parallel electric and magnetic fields \cite{Nielsen}. 
The physics of the chiral anomaly can be explained as follows. Consider the spectrum of
electrons in Weyl semimetal with only two Weyl points and in the presence of the magnetic field. Due to the linear dispersion around two Weyl points, two zero Landau levels are chiral and
velocities of electrons along the magnetic field have opposite sign for states around Weyl points with opposite chirality. The component of electric field parallel to the magnetic field will generate
charge imbalance between two chiral modes, pumping electrons between Weyl points with opposite chirality. The nonconservation of the chiral charge is called by the chiral anomaly.

Signatures of the chiral anomaly induced negative and quadratic in the magnetic field magnetoresistance, which possesses anisotropy as a function of the angle between the electric and magnetic fields \cite{SonSpivak, PhysRevLett.113.247203}, were observed in the theoretically proposed Dirac-Weyl semimetals: 
$\textrm{Cd}_3\textrm{As}_2$, $\textrm{TaAs}$, $\textrm{NbAs}$, $\mathrm{Na_3Bi}$, $\mathrm{ ZrTe_5}$ 
\cite{CdAs, bib:WSM1, bib:WSM2, Kharzeev, Xiong413, Felser}.

Although some attention was paid to the anomalous transport properties of Weyl semimetals,
to the best of our knowledge the interplay of the chiral anomaly and nonlinear optical properties is less explored \cite{Cort, Moore1}.
Stimulated by optical spectroscopy experiments of Weyl semimetals \cite{Weyl_optics1,Weyl_optics2}, 
we propose the theoretical study of the effect of the chiral anomaly on nonlinear optical phenomena in these systems. 

In this paper we study second harmonic generation (SHG) in centrosymmetric Weyl semimetal with broken time reversal symmetry. The SHG is the nonlinear optical effect related to the response of the electric current at a double frequency of the incident electromagnetic wave. The SHG in systems with spatial inversion centre in the lowest order is linear in the wave-vector of the electromagnetic field and quadratic in the electromagnetic field amplitudes. Neglecting nontrivial topology of the band structure of Weyl semimetals the theory of SHG in these systems can be obtained by generalizing the theory of SHG in graphene \cite{Misha} to three spatial dimensions. 

We show that the topologically nontrivial electronic structure of Weyl semimetals might be probed via SHG. Although, the effect of chiral anomaly is absent in Weyl semimetals radiated by the transverse electromagnetic wave, applying constant magnetic field to the system gives rise to anomaly contributions to SHG. 

We find that electromagnetic field with frequency $\omega$ and amplitude $(\mathbf{E}_0,\mathbf{B}_0)$ radiating two-valley Weyl semimetal subject to applied constant magnetic field $\mathbf{B}_c$ gives rise to SHG contribution in the form
\begin{equation}\label{main_answer}
\mathbf{J}^{(2)} =  \frac{\sigma}{1 -i\omega \tau_v/2}\frac{\tau_v}{2 \tau}\left(\frac{\omega_c}{\mu}\right)^2\frac{(\mathbf{E}_0\cdot \mathbf{B}_c)}{B_c^2}\mathbf{B}_0,
\end{equation}
where $\mu$ is the Fermi energy, $\omega_c=-ev^2B_c/c\mu$ is the cyclotron frequency, $e<0$ is the electron charge, $v$ is the Fermi velocity, $1/\tau=1/\tau_0+1/\tau_v$ is the relaxation rate, in which $\tau_{0(v)}$ defines intra (inter) valley scattering time, $\sigma = 2e^2\nu D $ is the conductivity, $\nu = \mu^2/2\pi^2 v^3$ is the electronic density of states per spin and per valley, and $D = v^2 \tau/3$ is the diffusion coefficient.

Electromagnetic wave generates ac current at a double frequency along constant magnetic field $\mathbf{B}_c$ only if there exists a component of wave-vector $\mathbf{q}$, which is transverse to this field. SHG has a peak at frequency $\omega\sim 0$ with a half-width $2/\tau_v$. It is also shown that cyclotron contributions to $\mathbf{J}^{(2)}$, which are explicitly written in Supplemental Material (SM), have peaks at frequencies $\omega_c$ and $\omega_c/2$ with half-widths $1/\tau$.

\emph{Model}. We consider a minimal model of centrosymmetric Weyl semimetal with only two valleys in its band structure. Conduction and valence bands in each valley touch at a single Weyl point in momentum space. Low energy excitations around these two Weyl points of opposite chirality, at the same energy, and separated in the momentum space are described by the Hamiltonian
\begin{equation}
H_{s}=s v \bsigma \cdot(-i\mathbf{\nabla}+sQ\hat{e}_z)-\mu,
\end{equation}
where $s=\pm$ defines chirality of a given valley, $2Q$ is the separation of two Weyl points along $z$-axis in momentum space, $\hat{e}_z$ is the unit vector ($\hbar =1$ throughout our calculations). We consider that the Fermi level crosses conduction band only, namely we set $\mu>0$. 

The SHG in Weyl semimetal is studied in the following way. The system is placed in the constant magnetic field $\mathbf{B}_c$, which is directed along $z$-axis, and is radiated by the transverse electromagnetic wave. The electric $\mathbf{E}(\mathbf{r},t)$ and magnetic $\mathbf{B}(\mathbf{r},t)$ fields are taken in the form
\begin{subequations}\label{EB}
\begin{align} 
&\mathbf{E}(\mathbf{r},t) =  \mathbf{E}_0 e^{i (\mathbf{qr} -\omega t)}  +\textrm{c.c.},\\
&\mathbf{B}(\mathbf{r},t) =  \mathbf{B}_c+[\mathbf{B}_0 e^{i (\mathbf{qr} -\omega t)}  +\textrm{c.c.}],
\end{align} 
\end{subequations}
where $\mathbf{E}_0=(E_{0,x},E_{0,y},E_{0,z})$ and $ \mathbf{B}_0=(B_{0,x},B_{0,y},B_{0,z})$ are the amplitudes of the incident transverse electromagnetic field with wave-vector $\mathbf{q}$.
Here it is assumed that in the medium $\mathbf{E}_0\cdot\mathbf{B}_0 =0$, while generally $\mathbf{E}_0\cdot\mathbf{B}_c \neq 0$.

Assuming that the cyclotron frequency related to magnetic field $\mathbf{B}_c$ is much smaller than the Fermi energy as well as the frequency and wave-vector of incident electromagnetic wave are much smaller then the Fermi energy and Fermi momentum respectively, $\omega\ll\mu$ and $q\ll \mu/v$, the SHG can be studied within the framework of Boltzmann equation.

\emph{Kinetic equation}. We proceed further by writing down kinetic equation for the distribution function of electrons $n_{\mathbf{k},s}(\mathbf{r}, t)$ in the phase space $(\mathbf{k},\mathbf{r})$ for the valley s of Weyl semimetal in the presence of electric  $\mathbf{E}(\mathbf{r}, t)$ and magnetic $\mathbf{B}(\mathbf{r}, t)$ fields ~\cite{PhysRevD.87.085016, RevModPhys.82.1959} 
\begin{eqnarray}\label{kinur}
\frac{\partial n_{\mathbf{k},s}}{\partial t} + \dot{\mathbf{k}}\cdot\frac{\partial n_{\mathbf{k},s}}{\partial \mathbf{k}} +\dot{\mathbf{r}}\cdot\frac{\partial n_{\mathbf{k},s}}{\partial \mathbf{r}} = \mathcal{I}_{\mathbf{k},s}.
\end{eqnarray} 
Kinetic equation is supplemented by the semiclassical equations of motion, which include effects of the Berry curvature and orbital magnetic moment of electrons
\begin{subequations}\label{EqMotion1}
\begin{align}
\dot{\mathbf{k}} &= e\tilde{\mathbf{E}} +\frac{e}{c}\dot{\mathbf{r}}\times\mathbf{B},\\
\dot{\mathbf{r}} &= \mathbf{v}_s +\dot{\mathbf{k}}\times\mathbf{\Omega}_{\mathbf{k},s},
\end{align}
\end{subequations} 
where $\mathbf{v}_s = \frac{\partial \mathcal{E}_{\mathbf{k},s}}{\partial \mathbf{k}}$ is the group velocity of the wave packet, which energy $\mathcal{E}_{\mathbf{k},s}$ includes correction due to magnetic orbital moment $\mathbf{m}_{\mathbf{k},s}=-e v k \mathbf{\Omega}_{\mathbf{k},s}$ of chiral fermions \cite{PhysRevD.87.085016,RevModPhys.82.1959}
\begin{equation}\label{energy_mod}
\mathcal{E}_{\mathbf{k},s} = v k\left(1 - \frac{e}{c}\mathbf{B}\cdot\mathbf{\Omega}_{\mathbf{k},s}\right),
\end{equation}
and $e\tilde{\mathbf{E}} = e\mathbf{E}-\partial \mathcal{E}_{\mathbf{k},s}/\partial \mathbf{r}$. Thus, the velocity of the wave packet is given by $\mathbf{v}_s=v\hat{\mathbf{k}}[1+\frac{2e}{c}\mathbf{B}\cdot\mathbf{\Omega}_{\mathbf{k},s}] -  \frac{sve}{2ck^2}\mathbf{B}$.

Nontrivial topology of the band structure of Weyl semimetal gives rise to the Berry curvature, which is for the s-th valley given by \cite{PhysRevD.87.085016}
$
\mathbf{\Omega}_{\mathbf{k},s} = s\hat{\mathbf{k}}/2k^2,
$
where $k\equiv|\mathbf{k}|$ and $\hat{\mathbf{k}}\equiv\mathbf{k}/k$ is the unit vector. Surface integral of the Berry curvature around the Weyl point with chirality $s=\pm $ satisfies $\int_S d\mathbf{ S}\cdot\mathbf{\Omega}_{\mathbf{k}, s}=2\pi s$.

Solutions to equations of motion Eq. \ref{EqMotion1} are given by
\begin{subequations}\label{EqMotion2}
\begin{align}
\dot{\mathbf{r}} &= D^{-1}_{\mathbf{k},s}\left[ \mathbf{v}_s + e\tilde{\mathbf{E}}\times\mathbf{\Omega}_{\mathbf{k},s} + \frac{e}{c}(\mathbf{v}_s\cdot\mathbf{\Omega}_{\mathbf{k}, s})\mathbf{B}\right],\\
\dot{\mathbf{k}} &= D^{-1}_{\mathbf{k},s}\left[ e\tilde{\mathbf{E}}+ \frac{e}{c} \mathbf{v}_s\times \mathbf{B} + \frac{e^2}{c}( \tilde{\mathbf{E}}\cdot \mathbf{B})\mathbf{\Omega}_{\mathbf{k},s}\right],\\
D&_{\mathbf{k},s}=1+\frac{e}{c}\mathbf{B}\cdot\mathbf{\Omega}_{\mathbf{k}, s}.
\end{align}
\end{subequations} 
Coupling of the electromagnetic field with the Berry curvature and magnetic moment of chiral fermions gives rise to the unusual properties of Weyl semimetals \cite{Anton_review}. In particular, second and third terms in $\dot{\mathbf{r}}$ describes the anomalous Hall and chiral magnetic effects respectively, third term in $\dot{\mathbf{k}} $ describes the chiral anomaly within the quasiclassical approach.

Collision integral in the kinetic equation Eq. \ref{kinur} is assumed in the relaxation time approximation \cite{RevModPhys.82.1959, Vova_Weyl}
\begin{equation}
\mathcal{I}_{\mathbf{k},\pm}  = \frac{\langle n_{\mathbf{k},\pm}\rangle - n_{\mathbf{k},\pm}}{\tau_0} + \frac{\langle n_{\mathbf{k},\mp}\rangle - n_{\mathbf{k},\pm}}{\tau_v},
\end{equation}
where $\tau_0$ is the intra-valley scattering time, $\tau_v$
is the inter-valley scattering time, and triangle brackets $\langle ... \rangle$ mean integration over the directions of momentum $\mathbf{k}$ taking into account the change of the phase space in the presence of the magnetic field
\begin{equation}
\langle n_{\mathbf{k},s}\rangle \equiv \int \frac{d \vec{\Theta}}{4\pi} D_{\mathbf{k},s} n_{\mathbf{k},s},
\end{equation}
where $\frac{d^3k}{(2\pi)^3} \equiv \frac{k^2dk}{2\pi^2}\frac{d \vec{\Theta}}{4\pi}$.
It is convenient to define a function \cite{Vova_Weyl}
$
\Lambda=  \langle n_{\mathbf{k},+}\rangle-\langle n_{\mathbf{k},-} \rangle
$ and rewrite collision integral in the form
\begin{equation}\label{stolk}
\mathcal{I}_{\mathbf{k},s}  = \frac{\langle n_{\mathbf{k},s}\rangle - n_{\mathbf{k},s}}{\tau} - \frac{s}{\tau_v}\Lambda,
\end{equation}
where $1/\tau = 1/\tau_0 + 1/\tau_v$ is the relaxation rate. Kinetic equations for the distribution functions in different valleys are coupled to each other by the last term in Eq. \ref{stolk}, which is proportional to the inter-valley relaxation rate, $1/\tau_v$.

We note that relaxation times generally depend on the Berry curvature in the presence of the magnetic field
$\tau_{0,v} \rightarrow \tau_{0,v} + \tilde{\tau}_{0,v}\frac{e}{c}\mathbf{B}\cdot\Omega_{\mathbf{k},s}$, where $\tau_{0,v}$ and $ \tilde{\tau}_{0,v}$ are some functions, which depend on the probabilities of elastic scattering of electrons on disorder \cite{RevModPhys.82.1959}. In what follows, we assume that relaxation times are independent from energy and magnetic field.

Finally, expression for the current density $\mathbf{J}(\mathbf{r}, t)$ is defined as \cite{RevModPhys.82.1959}
\begin{equation}\label{maincurrent}
\mathbf{J} = \sum_{s=\pm} \int \frac{d^3k}{(2\pi)^3} \bigg\{ eD_{\mathbf{k},s}\dot{\mathbf{r}}~n_{\mathbf{k},s}+\frac{\partial }{\partial \mathbf{r}}\times (\mathbf{m}_{\mathbf{k},s}n_{\mathbf{k},s} )\bigg\},
\end{equation}
where first term originates from the group velocity of the wavepacket, which contains contributions from the Berry curvature and magnetic orbital moment, while second term is coming from the curl of orbital magnetization of electrons. 

Inserting expression from Eq.~\ref{EqMotion2}a into Eq.~\ref{maincurrent} one obtains
\begin{eqnarray}\nonumber\label{Current11}
\mathbf{J} &=& e\sum_{s=\pm} \int \frac{d^3k}{(2\pi)^3} \bigg\{ \bigg[ v\hat{\mathbf{k}}(1+\frac{2e}{c}\mathbf{B}\cdot\mathbf{\Omega}_{\mathbf{k},s}) + e\tilde{\mathbf{E}}\times\mathbf{\Omega}_{\mathbf{k},s} 
\\
&+& \frac{sve^2}{2c^2k^2}\mathbf{B}(\mathbf{B}\cdot\mathbf{\Omega}_{\mathbf{k},s}) \bigg]  n_{\mathbf{k},s}
-vk \frac{\partial n_{\mathbf{k},s}}{\partial \mathbf{r}}\times\mathbf{\Omega}_{\mathbf{k},s}\bigg\}.
\end{eqnarray}
Here it is useful to note that terms in the integrand are proportional to either momentum $\mathbf{k}$, chirality number $s$, or their product $\mathbf{k}s$. 

\emph{Second Harmonic Generation}. SHG is a response of the electric current at double frequency~$2\omega$ of propagating electromagnetic field with frequency~$\omega$. This effect is absent in
systems with inversion centre provided the distribution function in the presence of the electromagnetic field is spatial independent. Although second harmonic can be generated at the surface of the system \cite{Misha}, we focus on the bulk properties of this effect and neglect surface contribution.

Let us remind the region of applicability of the quasiclassical approximation, which is used in this paper. We consider the wave-vector and the frequency of the electromagnetic wave to be much smaller than the Fermi momentum and Fermi energy of electrons $q\ll \mu/v,~\omega\ll \mu$. Thus, we do not consider interband contributions to SHG, which become important at $\omega\sim 2\mu$. 
The magnetic field $\mathbf{B}_c$ is assumed to satisfy
\begin{equation}
1 \lesssim \omega_c\tau \ll \mu\tau.
\end{equation}
Classical effects of the magnetic field become important when the cyclotron frequency $\omega_c$ is larger than the 
electron inverse mean free time $1/\tau$.

We search for the approximate solution of kinetic equation Eq. \ref{kinur}, keeping contributions to the distribution function up to second power of incident electromagnetic field
\begin{equation} \label{expansion}
n_{\mathbf{k},s}=n^{(0)}_{\mathbf{k},s} + [n^{(1)}_{\mathbf{k},s} e^{i(\mathbf{q}\cdot\mathbf{r}-\omega t)}+n^{(2)}_{\mathbf{k},s}e^{2i(\mathbf{q}\cdot\mathbf{r}-\omega t)} + \textrm{c.c.}],
\end{equation}
where 
$n^{(0)}_{\mathbf{k},s} = [1+e^{\left(v k(1 - \frac{e}{c}\mathbf{B}_c\cdot\mathbf{\Omega}_{\mathbf{k},s}) -\mu\right)/T}]^{-1}$
is the Fermi distribution function defined with dispersion relation Eq. \ref{energy_mod}, which is modified by the orbital magnetic moment in the presence of magnetic field $\mathbf{B}_c$, and $T\ll \mu$ is the temperature.
We similarly expand the current density 
$
\mathbf{J}=\mathbf{J}^{(0)}+[\mathbf{J}^{(1)}e^{i(\mathbf{q}\cdot\mathbf{r}-\omega t)}+\mathbf{J}^{(2)}e^{2i(\mathbf{q}\cdot\mathbf{r}-\omega t)}+\textrm{c.c.}]$ and intervalley coupling function 
$
\Lambda=\Lambda^{(0)} +[\Lambda^{(1)} e^{i(\mathbf{q}\cdot\mathbf{r}-\omega t)}+\Lambda^{(2)}e^{2i(\mathbf{q}\cdot\mathbf{r}-\omega t)} + \textrm{c.c.}]
$. 

Absorbing contributions up to the second power in the amplitude of incident radiation,
one obtains expression for the current density $\mathbf{J}^{(2)}$ at double frequency of electromagnetic wave in the form
\begin{eqnarray}\nonumber\label{Current_total}
&\mathbf{J}^{(2)}& = e\sum_{s=\pm} \int \frac{d^3k}{(2\pi)^3} \bigg\{\bigg[v\hat{\mathbf{k}}\frac{2e}{c}(\mathbf{B}_0\cdot\mathbf{\Omega}_{\mathbf{k},s}) +e\mathbf{E}_0\times\mathbf{\Omega}_{\mathbf{k},s}
\\\nonumber
&+&  \frac{sve^2}{2c^2k^2}[(\mathbf{B}_c\cdot\mathbf{\Omega}_{\mathbf{k},s})\mathbf{B}_0+(\mathbf{B}_0\cdot\mathbf{\Omega}_{\mathbf{k},s})\mathbf{B}_c]
\\\nonumber
&+& ivk\frac{e}{c}(\mathbf{B}_0\cdot\mathbf{\Omega}_{\mathbf{k},s})[\mathbf{q}\times \mathbf{\Omega}_{\mathbf{k},s}]\bigg]n^{(1)}_{\mathbf{k},s}-2 i vk [\mathbf{q} \times\mathbf{\Omega}_{\mathbf{k},s} ]n^{(2)}_{\mathbf{k},s}
\\
&+&  v\hat{\mathbf{k}}(1+\frac{2e}{c}\mathbf{B}_c\cdot\mathbf{\Omega}_{\mathbf{k},s})  n^{(2)}_{\mathbf{k},s} \bigg\}.
\end{eqnarray}

We solve kinetic equation including second order corrections in amplitude of the electromagnetic field to the distribution function assuming the long wave limit of the response $\omega\gg vq$, and in first order in $\frac{2e}{c}\mathbf{B}_c\cdot\mathbf{\Omega}_{\mathbf{k},s}$, see SM. 
Therefore we solve spatially independent kinetic equation in order to capture the correction to $EB$ mechanism \cite{Misha} of SHG originating from the topologically nontrivial electronic band structure. Wave-vector dependent contribution to SHG, the so-called $qE^2$ mechanism, is not under the scope of the present paper. Correction to $qE^2$ mechanism is a perturbation with a small parameter $vq\tau \frac{\mu}{\omega_c} \ll 1$, which is satisfied for $\omega\ll \omega_c$. We note that $q$-dependent corrections become important for frequencies $\omega\sim \omega_c$, at which $\frac{v}{c}\mu\tau$ might be of the order of unity. 

It is practical to use cylindrical coordinate representation, which simplifies solution of the kinetic equation in the presence of the magnetic field. Let us introduce new variables $(k_{\perp}\cos(\phi), k_{\perp}\sin(\phi), k_z) = (k_x, k_y, k_z)$, which allows to rewrite 
$
\frac{ev}{c}[\hat{\mathbf{k}}\times\mathbf{B}_c]\cdot \frac{\partial n}{\partial \mathbf{k}} = -\frac{ev}{ck}B_c \frac{\partial n}{\partial \phi}
$. 

We obtain kinetic equation for the first order correction $n^{(1)}_{\mathbf{k},s}$ in the form
\begin{eqnarray}\nonumber\label{n11}
&\bigg[&-i\omega n^{(1)}_{\mathbf{k},s} + \frac{s\Lambda^{(1)}}{\tau_v} +\frac{n^{(1)}_{\mathbf{k},s}-\langle n^{(1)}_{\mathbf{k},s}\rangle}{\tau}\bigg]\frac{D_{\mathbf{k},s}}{\tilde{D}_{\mathbf{k},s}}
-\frac{ev}{ck}B_c\frac{\partial  n^{(1)}_{\mathbf{k},s}}{\partial\phi}
\\
&=&\frac{-e}{\tilde{D}_{\mathbf{k},s}}\bigg[\mathbf{E}_0\cdot\hat{\mathbf{k}} +\frac{se}{2ck^2} (\mathbf{E}_0+\frac{v}{c}\mathbf{B}_0\times\hat{\mathbf{k}}_{\perp})\cdot\mathbf{B}_c\bigg]\frac{\partial  n^{(0)}_{\mathbf{k},s}}{\partial k},~~~~~
\end{eqnarray}
where $\mathbf{E}_0\cdot \hat{\mathbf{k}} = E_{0,z} \hat{k}_z+ [E_{0,x} \cos(\phi)+E_{0,y} \sin(\phi)]\hat{k}_{\perp}$ and $\tilde{D}_{\mathbf{k},s}=2D_{\mathbf{k},s}-1$ is introduced for brevity. We observe that the right hand side of Eq. \ref{n11} contains terms that describe chiral anomaly and orbital magnetic moment contributions to SHG. Solution of the first order differential Eq. \ref{n11} is given by expression in Eq. \ref{n1_Solution} in SM.

Using the parity symmetry of the integrand in Eq. \ref{Current11} with respect to $s\mathbf{k}$, we keep terms that give non vanishing contribution to the second harmonic generation and neglect all other terms in equation for the second order correction to distribution function
\begin{eqnarray}\nonumber\label{n22}
\bigg[&-&2i\omega+\frac{1}{\tau}-\frac{ev}{ck}B_cD_{\mathbf{k},s}\frac{\partial}{\partial \phi}\bigg]n_{\mathbf{k},s}^{(2)}=-\frac{ev}{c}[\hat{\mathbf{k}}\times\mathbf{B}_0]\cdot \frac{\partial n_{\mathbf{k},s}^{(1)}}{\partial \mathbf{k}}\\
&+&
\frac{s e^2 \mathbf{B}_0\cdot\hat{\mathbf{k}}}{2ck^2D^2_{\mathbf{k},s}}
\bigg[\mathbf{E}_0\cdot\hat{\mathbf{k}} +\frac{se}{2ck^2} \mathbf{E}_0\cdot\mathbf{B}_c\bigg]\frac{\partial n^{(0)}_{\mathbf{k},s}}{\partial k}
\end{eqnarray}  
First term on the rhs of this equation describes $EB$-mechanism of the SHG, while other terms are corrections to this mechanism due to the topological band structure of Weyl semimetal. 
We also observe that correction to the inter-valley coupling $\Lambda^{(2)}$ does not contribute to SHG, which means that nontrivial topological corrections to $n_{\mathbf{k},s}^{(2)}$ can exist even in the absence of the inter-valley relaxation. Although right hand side of Eq. \ref{n22} still contains terms $\propto \cos(2\phi)$ and $\propto \sin(2\phi)$, they will also not contribute to $\mathbf{J}^{(2)}$. 
Therefore, solution of Eq. \ref{n22} reduces to the solution of Eq. \ref{n11} and is given by Eq. \ref{n2_appendix} in SM.  Quite lengthly general expression for SHG is given in SM by summarizing contributions from $n_{\mathbf{k},s}^{(1,2)}$, Eqs. \ref{Cur_split_appendix}, \ref{Current_n1}, and \ref{Cur_n2_appendix}.

Let us now discuss special cases of the solution. We find that SHG vanishes in the limit when $B_c\neq0$ and external field has only electric component, $B_0=0$, $E_0\neq0$. This result means that there is no contribution to $\mathbf{J}^{(2)}$, which is $\propto B_c E_{0,i}E_{0,j}$, where $i,j\in {x,y,z}$. In the 
limit when external magnetic field is turned off, $B_c=0$, while $E_0,B_0\neq 0$, SHG is determined by the well known $EB$-mechanism, $\mathbf{J}^{(2)} \propto [\mathbf{B}_0\times\mathbf{E}_0]$.
In this paper we are interested in the case when the chiral anomaly is the dominant contribution to SHG. Assuming the inter-valley scattering time $\tau_v$ to be much longer than the intra-valley scattering time $\tau_0$, $\tau_v\gg \tau_0$, we arrive at our main result, the effect of the chiral anomaly on SHG, given in Eq. \ref{main_answer}. 

\emph{Summary}. In conclusion, we have investigated the effect of chiral anomaly on SHG in centrosymmetric Weyl semimetal within the kinetic equation approach. 
We remind that SHG in systems with inversion centre requires incident radiation with finite wave-vector, 
while propagating electromagnetic wave with transverse polarization can not lead to the chiral anomaly. 
We show that applying a constant magnetic field in addition to propagating electromagnetic wave gives rise to the observable contribution of the chiral anomaly to SHG.

\emph{Acknowledgements}. 
We would like to acknowledge discussions with Mikhail Glazov and Vladimir Zyuzin. AAZ was financially supported by the Swedish Research Council Grant No. 642-2013-7837, Swiss SNF, and the NCCR Quantum Science and Technology. AYZ acknowledges support from the Russian Scientific Fund No. 16-42-01067.

\bibliography{Bib2ndHarm}{}

\begin{thebibliography}{29}%
\makeatletter
\providecommand \@ifxundefined [1]{%
 \@ifx{#1\undefined}
}%
\providecommand \@ifnum [1]{%
 \ifnum #1\expandafter \@firstoftwo
 \else \expandafter \@secondoftwo
 \fi
}%
\providecommand \@ifx [1]{%
 \ifx #1\expandafter \@firstoftwo
 \else \expandafter \@secondoftwo
 \fi
}%
\providecommand \natexlab [1]{#1}%
\providecommand \enquote  [1]{``#1''}%
\providecommand \bibnamefont  [1]{#1}%
\providecommand \bibfnamefont [1]{#1}%
\providecommand \citenamefont [1]{#1}%
\providecommand \href@noop [0]{\@secondoftwo}%
\providecommand \href [0]{\begingroup \@sanitize@url \@href}%
\providecommand \@href[1]{\@@startlink{#1}\@@href}%
\providecommand \@@href[1]{\endgroup#1\@@endlink}%
\providecommand \@sanitize@url [0]{\catcode `\\12\catcode `\$12\catcode
  `\&12\catcode `\#12\catcode `\^12\catcode `\_12\catcode `\%12\relax}%
\providecommand \@@startlink[1]{}%
\providecommand \@@endlink[0]{}%
\providecommand \url  [0]{\begingroup\@sanitize@url \@url }%
\providecommand \@url [1]{\endgroup\@href {#1}{\urlprefix }}%
\providecommand \urlprefix  [0]{URL }%
\providecommand \Eprint [0]{\href }%
\providecommand \doibase [0]{http://dx.doi.org/}%
\providecommand \selectlanguage [0]{\@gobble}%
\providecommand \bibinfo  [0]{\@secondoftwo}%
\providecommand \bibfield  [0]{\@secondoftwo}%
\providecommand \translation [1]{[#1]}%
\providecommand \BibitemOpen [0]{}%
\providecommand \bibitemStop [0]{}%
\providecommand \bibitemNoStop [0]{.\EOS\space}%
\providecommand \EOS [0]{\spacefactor3000\relax}%
\providecommand \BibitemShut  [1]{\csname bibitem#1\endcsname}%
\let\auto@bib@innerbib\@empty
\bibitem [{\citenamefont {Wan}\ \emph {et~al.}(2011)\citenamefont {Wan},
  \citenamefont {Turner}, \citenamefont {Vishwanath},\ and\ \citenamefont
  {Savrasov}}]{Weyl}%
  \BibitemOpen
  \bibfield  {author} {\bibinfo {author} {\bibfnamefont {X.}~\bibnamefont
  {Wan}}, \bibinfo {author} {\bibfnamefont {A.~M.}\ \bibnamefont {Turner}},
  \bibinfo {author} {\bibfnamefont {A.}~\bibnamefont {Vishwanath}}, \ and\
  \bibinfo {author} {\bibfnamefont {S.~Y.}\ \bibnamefont {Savrasov}},\ }\href
  {\doibase 10.1103/PhysRevB.83.205101} {\bibfield  {journal} {\bibinfo
  {journal} {Phys. Rev. B}\ }\textbf {\bibinfo {volume} {83}},\ \bibinfo
  {pages} {205101} (\bibinfo {year} {2011})}\BibitemShut {NoStop}%
\bibitem [{\citenamefont {Yang}\ \emph {et~al.}(2011)\citenamefont {Yang},
  \citenamefont {Lu},\ and\ \citenamefont {Ran}}]{Weyl1}%
  \BibitemOpen
  \bibfield  {author} {\bibinfo {author} {\bibfnamefont {K.-Y.}\ \bibnamefont
  {Yang}}, \bibinfo {author} {\bibfnamefont {Y.-M.}\ \bibnamefont {Lu}}, \ and\
  \bibinfo {author} {\bibfnamefont {Y.}~\bibnamefont {Ran}},\ }\href {\doibase
  10.1103/PhysRevB.84.075129} {\bibfield  {journal} {\bibinfo  {journal} {Phys.
  Rev. B}\ }\textbf {\bibinfo {volume} {84}},\ \bibinfo {pages} {075129}
  (\bibinfo {year} {2011})}\BibitemShut {NoStop}%
\bibitem [{\citenamefont {Burkov}\ and\ \citenamefont {Balents}(2011)}]{Weyl2}%
  \BibitemOpen
  \bibfield  {author} {\bibinfo {author} {\bibfnamefont {A.~A.}\ \bibnamefont
  {Burkov}}\ and\ \bibinfo {author} {\bibfnamefont {L.}~\bibnamefont
  {Balents}},\ }\href {\doibase 10.1103/PhysRevLett.107.127205} {\bibfield
  {journal} {\bibinfo  {journal} {Phys. Rev. Lett.}\ }\textbf {\bibinfo
  {volume} {107}},\ \bibinfo {pages} {127205} (\bibinfo {year}
  {2011})}\BibitemShut {NoStop}%
\bibitem [{\citenamefont {Xu}\ \emph {et~al.}(2011)\citenamefont {Xu},
  \citenamefont {Weng}, \citenamefont {Wang}, \citenamefont {Dai},\ and\
  \citenamefont {Fang}}]{Weyl3}%
  \BibitemOpen
  \bibfield  {author} {\bibinfo {author} {\bibfnamefont {G.}~\bibnamefont
  {Xu}}, \bibinfo {author} {\bibfnamefont {H.}~\bibnamefont {Weng}}, \bibinfo
  {author} {\bibfnamefont {Z.}~\bibnamefont {Wang}}, \bibinfo {author}
  {\bibfnamefont {X.}~\bibnamefont {Dai}}, \ and\ \bibinfo {author}
  {\bibfnamefont {Z.}~\bibnamefont {Fang}},\ }\href {\doibase
  10.1103/PhysRevLett.107.186806} {\bibfield  {journal} {\bibinfo  {journal}
  {Phys. Rev. Lett.}\ }\textbf {\bibinfo {volume} {107}},\ \bibinfo {pages}
  {186806} (\bibinfo {year} {2011})}\BibitemShut {NoStop}%
\bibitem [{\citenamefont {Murakami}(2007)}]{Murakami}%
  \BibitemOpen
  \bibfield  {author} {\bibinfo {author} {\bibfnamefont {S.}~\bibnamefont
  {Murakami}},\ }\href {http://stacks.iop.org/1367-2630/9/i=9/a=356} {\bibfield
   {journal} {\bibinfo  {journal} {New Journal of Physics}\ }\textbf {\bibinfo
  {volume} {9}},\ \bibinfo {pages} {356} (\bibinfo {year} {2007})}\BibitemShut
  {NoStop}%
\bibitem [{\citenamefont {Xu}\ \emph {et~al.}(2015)\citenamefont {Xu},
  \citenamefont {Belopolski}, \citenamefont {Alidoust}, \citenamefont
  {Neupane}, \citenamefont {Bian}, \citenamefont {Zhang}, \citenamefont
  {Sankar}, \citenamefont {Chang}, \citenamefont {Yuan}, \citenamefont {Lee},
  \citenamefont {Huang}, \citenamefont {Zheng}, \citenamefont {Ma},
  \citenamefont {Sanchez}, \citenamefont {Wang}, \citenamefont {Bansil},
  \citenamefont {Chou}, \citenamefont {Shibayev}, \citenamefont {Lin},
  \citenamefont {Jia},\ and\ \citenamefont {Hasan}}]{Xu613}%
  \BibitemOpen
  \bibfield  {author} {\bibinfo {author} {\bibfnamefont {S.-Y.}\ \bibnamefont
  {Xu}}, \bibinfo {author} {\bibfnamefont {I.}~\bibnamefont {Belopolski}},
  \bibinfo {author} {\bibfnamefont {N.}~\bibnamefont {Alidoust}}, \bibinfo
  {author} {\bibfnamefont {M.}~\bibnamefont {Neupane}}, \bibinfo {author}
  {\bibfnamefont {G.}~\bibnamefont {Bian}}, \bibinfo {author} {\bibfnamefont
  {C.}~\bibnamefont {Zhang}}, \bibinfo {author} {\bibfnamefont
  {R.}~\bibnamefont {Sankar}}, \bibinfo {author} {\bibfnamefont
  {G.}~\bibnamefont {Chang}}, \bibinfo {author} {\bibfnamefont
  {Z.}~\bibnamefont {Yuan}}, \bibinfo {author} {\bibfnamefont {C.-C.}\
  \bibnamefont {Lee}}, \bibinfo {author} {\bibfnamefont {S.-M.}\ \bibnamefont
  {Huang}}, \bibinfo {author} {\bibfnamefont {H.}~\bibnamefont {Zheng}},
  \bibinfo {author} {\bibfnamefont {J.}~\bibnamefont {Ma}}, \bibinfo {author}
  {\bibfnamefont {D.~S.}\ \bibnamefont {Sanchez}}, \bibinfo {author}
  {\bibfnamefont {B.}~\bibnamefont {Wang}}, \bibinfo {author} {\bibfnamefont
  {A.}~\bibnamefont {Bansil}}, \bibinfo {author} {\bibfnamefont
  {F.}~\bibnamefont {Chou}}, \bibinfo {author} {\bibfnamefont {P.~P.}\
  \bibnamefont {Shibayev}}, \bibinfo {author} {\bibfnamefont {H.}~\bibnamefont
  {Lin}}, \bibinfo {author} {\bibfnamefont {S.}~\bibnamefont {Jia}}, \ and\
  \bibinfo {author} {\bibfnamefont {M.~Z.}\ \bibnamefont {Hasan}},\ }\href
  {\doibase 10.1126/science.aaa9297} {\bibfield  {journal} {\bibinfo  {journal}
  {Science}\ }\textbf {\bibinfo {volume} {349}},\ \bibinfo {pages} {613}
  (\bibinfo {year} {2015})}\BibitemShut {NoStop}%
\bibitem [{\citenamefont {Lv}\ \emph {et~al.}(2015)\citenamefont {Lv},
  \citenamefont {Weng}, \citenamefont {Fu}, \citenamefont {Wang}, \citenamefont
  {Miao}, \citenamefont {Ma}, \citenamefont {Richard}, \citenamefont {Huang},
  \citenamefont {Zhao}, \citenamefont {Chen}, \citenamefont {Fang},
  \citenamefont {Dai}, \citenamefont {Qian},\ and\ \citenamefont
  {Ding}}]{PhysRevX.5.031013}%
  \BibitemOpen
  \bibfield  {author} {\bibinfo {author} {\bibfnamefont {B.~Q.}\ \bibnamefont
  {Lv}}, \bibinfo {author} {\bibfnamefont {H.~M.}\ \bibnamefont {Weng}},
  \bibinfo {author} {\bibfnamefont {B.~B.}\ \bibnamefont {Fu}}, \bibinfo
  {author} {\bibfnamefont {X.~P.}\ \bibnamefont {Wang}}, \bibinfo {author}
  {\bibfnamefont {H.}~\bibnamefont {Miao}}, \bibinfo {author} {\bibfnamefont
  {J.}~\bibnamefont {Ma}}, \bibinfo {author} {\bibfnamefont {P.}~\bibnamefont
  {Richard}}, \bibinfo {author} {\bibfnamefont {X.~C.}\ \bibnamefont {Huang}},
  \bibinfo {author} {\bibfnamefont {L.~X.}\ \bibnamefont {Zhao}}, \bibinfo
  {author} {\bibfnamefont {G.~F.}\ \bibnamefont {Chen}}, \bibinfo {author}
  {\bibfnamefont {Z.}~\bibnamefont {Fang}}, \bibinfo {author} {\bibfnamefont
  {X.}~\bibnamefont {Dai}}, \bibinfo {author} {\bibfnamefont {T.}~\bibnamefont
  {Qian}}, \ and\ \bibinfo {author} {\bibfnamefont {H.}~\bibnamefont {Ding}},\
  }\href {\doibase 10.1103/PhysRevX.5.031013} {\bibfield  {journal} {\bibinfo
  {journal} {Phys. Rev. X}\ }\textbf {\bibinfo {volume} {5}},\ \bibinfo {pages}
  {031013} (\bibinfo {year} {2015})}\BibitemShut {NoStop}%
\bibitem [{\citenamefont {Neupane}\ \emph {et~al.}(2014)\citenamefont
  {Neupane}, \citenamefont {Xu}, \citenamefont {Sankar}, \citenamefont
  {Alidoust}, \citenamefont {Bian}, \citenamefont {Liu}, \citenamefont
  {Belopolski}, \citenamefont {Chang}, \citenamefont {Jeng}, \citenamefont
  {Lin}, \citenamefont {Bansil}, \citenamefont {Chou},\ and\ \citenamefont
  {Hasan}}]{CdAs}%
  \BibitemOpen
  \bibfield  {author} {\bibinfo {author} {\bibfnamefont {M.}~\bibnamefont
  {Neupane}}, \bibinfo {author} {\bibfnamefont {S.-Y.}\ \bibnamefont {Xu}},
  \bibinfo {author} {\bibfnamefont {R.}~\bibnamefont {Sankar}}, \bibinfo
  {author} {\bibfnamefont {N.}~\bibnamefont {Alidoust}}, \bibinfo {author}
  {\bibfnamefont {G.}~\bibnamefont {Bian}}, \bibinfo {author} {\bibfnamefont
  {C.}~\bibnamefont {Liu}}, \bibinfo {author} {\bibfnamefont {I.}~\bibnamefont
  {Belopolski}}, \bibinfo {author} {\bibfnamefont {T.-R.}\ \bibnamefont
  {Chang}}, \bibinfo {author} {\bibfnamefont {H.-T.}\ \bibnamefont {Jeng}},
  \bibinfo {author} {\bibfnamefont {H.}~\bibnamefont {Lin}}, \bibinfo {author}
  {\bibfnamefont {A.}~\bibnamefont {Bansil}}, \bibinfo {author} {\bibfnamefont
  {F.}~\bibnamefont {Chou}}, \ and\ \bibinfo {author} {\bibfnamefont {M.~Z.}\
  \bibnamefont {Hasan}},\ }\href {http://dx.doi.org/10.1038/ncomms4786}
  {\bibfield  {journal} {\bibinfo  {journal} {Nat Commun}\ }\textbf {\bibinfo
  {volume} {5}} (\bibinfo {year} {2014})}\BibitemShut {NoStop}%
\bibitem [{\citenamefont {Zhang}\ \emph {et~al.}()\citenamefont {Zhang},
  \citenamefont {Xu}, \citenamefont {Belopolski}, \citenamefont {Yuan},
  \citenamefont {Lin}, \citenamefont {Tong}, \citenamefont {Alidoust},
  \citenamefont {Lee}, \citenamefont {Huang}, \citenamefont {Lin},
  \citenamefont {Neupane}, \citenamefont {Sanchez}, \citenamefont {Zheng},
  \citenamefont {Bian}, \citenamefont {Wang}, \citenamefont {Zhang},
  \citenamefont {Neupert}, \citenamefont {Hasan},\ and\ \citenamefont
  {Jia}}]{bib:WSM1}%
  \BibitemOpen
  \bibfield  {author} {\bibinfo {author} {\bibfnamefont {C.}~\bibnamefont
  {Zhang}}, \bibinfo {author} {\bibfnamefont {S.-Y.}\ \bibnamefont {Xu}},
  \bibinfo {author} {\bibfnamefont {I.}~\bibnamefont {Belopolski}}, \bibinfo
  {author} {\bibfnamefont {Z.}~\bibnamefont {Yuan}}, \bibinfo {author}
  {\bibfnamefont {Z.}~\bibnamefont {Lin}}, \bibinfo {author} {\bibfnamefont
  {B.}~\bibnamefont {Tong}}, \bibinfo {author} {\bibfnamefont {N.}~\bibnamefont
  {Alidoust}}, \bibinfo {author} {\bibfnamefont {C.-C.}\ \bibnamefont {Lee}},
  \bibinfo {author} {\bibfnamefont {S.-M.}\ \bibnamefont {Huang}}, \bibinfo
  {author} {\bibfnamefont {H.}~\bibnamefont {Lin}}, \bibinfo {author}
  {\bibfnamefont {M.}~\bibnamefont {Neupane}}, \bibinfo {author} {\bibfnamefont
  {D.~S.}\ \bibnamefont {Sanchez}}, \bibinfo {author} {\bibfnamefont
  {H.}~\bibnamefont {Zheng}}, \bibinfo {author} {\bibfnamefont
  {G.}~\bibnamefont {Bian}}, \bibinfo {author} {\bibfnamefont {J.}~\bibnamefont
  {Wang}}, \bibinfo {author} {\bibfnamefont {C.}~\bibnamefont {Zhang}},
  \bibinfo {author} {\bibfnamefont {T.}~\bibnamefont {Neupert}}, \bibinfo
  {author} {\bibfnamefont {M.~Z.}\ \bibnamefont {Hasan}}, \ and\ \bibinfo
  {author} {\bibfnamefont {S.}~\bibnamefont {Jia}},\ }\href@noop {} {\bibinfo
  {journal} {arXiv:1503.02630}\ }\BibitemShut {NoStop}%
\bibitem [{\citenamefont {Huang}\ \emph {et~al.}(2015)\citenamefont {Huang},
  \citenamefont {Zhao}, \citenamefont {Long}, \citenamefont {Wang},
  \citenamefont {Chen}, \citenamefont {Yang}, \citenamefont {Liang},
  \citenamefont {Xue}, \citenamefont {Weng}, \citenamefont {Fang},
  \citenamefont {Dai},\ and\ \citenamefont {Chen}}]{bib:WSM2}%
  \BibitemOpen
\bibfield  {journal} {  }\bibfield  {author} {\bibinfo {author} {\bibfnamefont
  {X.}~\bibnamefont {Huang}}, \bibinfo {author} {\bibfnamefont
  {L.}~\bibnamefont {Zhao}}, \bibinfo {author} {\bibfnamefont {Y.}~\bibnamefont
  {Long}}, \bibinfo {author} {\bibfnamefont {P.}~\bibnamefont {Wang}}, \bibinfo
  {author} {\bibfnamefont {D.}~\bibnamefont {Chen}}, \bibinfo {author}
  {\bibfnamefont {Z.}~\bibnamefont {Yang}}, \bibinfo {author} {\bibfnamefont
  {H.}~\bibnamefont {Liang}}, \bibinfo {author} {\bibfnamefont
  {M.}~\bibnamefont {Xue}}, \bibinfo {author} {\bibfnamefont {H.}~\bibnamefont
  {Weng}}, \bibinfo {author} {\bibfnamefont {Z.}~\bibnamefont {Fang}}, \bibinfo
  {author} {\bibfnamefont {X.}~\bibnamefont {Dai}}, \ and\ \bibinfo {author}
  {\bibfnamefont {G.}~\bibnamefont {Chen}},\ }\href {\doibase
  10.1103/PhysRevX.5.031023} {\bibfield  {journal} {\bibinfo  {journal} {Phys.
  Rev. X}\ }\textbf {\bibinfo {volume} {5}},\ \bibinfo {pages} {031023}
  (\bibinfo {year} {2015})}\BibitemShut {NoStop}%
\bibitem [{\citenamefont {Li}\ \emph {et~al.}(2016)\citenamefont {Li},
  \citenamefont {Kharzeev}, \citenamefont {Zhang}, \citenamefont {Huang},
  \citenamefont {Pletikosic}, \citenamefont {Fedorov}, \citenamefont {Zhong},
  \citenamefont {Schneeloch}, \citenamefont {Gu},\ and\ \citenamefont
  {Valla}}]{Kharzeev}%
  \BibitemOpen
  \bibfield  {author} {\bibinfo {author} {\bibfnamefont {Q.}~\bibnamefont
  {Li}}, \bibinfo {author} {\bibfnamefont {D.~E.}\ \bibnamefont {Kharzeev}},
  \bibinfo {author} {\bibfnamefont {C.}~\bibnamefont {Zhang}}, \bibinfo
  {author} {\bibfnamefont {Y.}~\bibnamefont {Huang}}, \bibinfo {author}
  {\bibfnamefont {I.}~\bibnamefont {Pletikosic}}, \bibinfo {author}
  {\bibfnamefont {A.~V.}\ \bibnamefont {Fedorov}}, \bibinfo {author}
  {\bibfnamefont {R.~D.}\ \bibnamefont {Zhong}}, \bibinfo {author}
  {\bibfnamefont {J.~A.}\ \bibnamefont {Schneeloch}}, \bibinfo {author}
  {\bibfnamefont {G.~D.}\ \bibnamefont {Gu}}, \ and\ \bibinfo {author}
  {\bibfnamefont {T.}~\bibnamefont {Valla}},\ }\href
  {http://dx.doi.org/10.1038/nphys3648} {\bibfield  {journal} {\bibinfo
  {journal} {Nat Phys}\ }\textbf {\bibinfo {volume} {12}},\ \bibinfo {pages}
  {550} (\bibinfo {year} {2016})}\BibitemShut {NoStop}%
\bibitem [{\citenamefont {Volovik}(2003)}]{Volovik}%
  \BibitemOpen
  \bibfield  {author} {\bibinfo {author} {\bibfnamefont {G.~E.}\ \bibnamefont
  {Volovik}},\ }\href@noop {} {\emph {\bibinfo {title} {The Universe in a
  Helium Droplet}}}\ (\bibinfo  {publisher} {Oxford University Press, Oxford},\
  \bibinfo {year} {2003})\BibitemShut {NoStop}%
\bibitem [{\citenamefont {Wehling}\ \emph {et~al.}(2014)\citenamefont
  {Wehling}, \citenamefont {Black-Schaffer},\ and\ \citenamefont
  {Balatsky}}]{Balatsky_Dirac}%
  \BibitemOpen
  \bibfield  {author} {\bibinfo {author} {\bibfnamefont {T.}~\bibnamefont
  {Wehling}}, \bibinfo {author} {\bibfnamefont {A.}~\bibnamefont
  {Black-Schaffer}}, \ and\ \bibinfo {author} {\bibfnamefont {A.}~\bibnamefont
  {Balatsky}},\ }\href {\doibase 10.1080/00018732.2014.927109} {\bibfield
  {journal} {\bibinfo  {journal} {Advances in Physics}\ }\textbf {\bibinfo
  {volume} {63}},\ \bibinfo {pages} {1} (\bibinfo {year} {2014})}\BibitemShut
  {NoStop}%
\bibitem [{\citenamefont {Burkov}(2015)}]{Anton_review}%
  \BibitemOpen
  \bibfield  {author} {\bibinfo {author} {\bibfnamefont {A.}~\bibnamefont
  {Burkov}},\ }\href@noop {} {\bibfield  {journal} {\bibinfo  {journal} {J.
  Phys.: Condens. Matter}\ }\textbf {\bibinfo {volume} {27}},\ \bibinfo {pages}
  {113201} (\bibinfo {year} {2015})}\BibitemShut {NoStop}%
\bibitem [{\citenamefont {Adler}(1969)}]{PhysRev.177.2426}%
  \BibitemOpen
  \bibfield  {author} {\bibinfo {author} {\bibfnamefont {S.~L.}\ \bibnamefont
  {Adler}},\ }\href {\doibase 10.1103/PhysRev.177.2426} {\bibfield  {journal}
  {\bibinfo  {journal} {Phys. Rev.}\ }\textbf {\bibinfo {volume} {177}},\
  \bibinfo {pages} {2426} (\bibinfo {year} {1969})}\BibitemShut {NoStop}%
\bibitem [{\citenamefont {Bell}\ and\ \citenamefont {Jackiw}(1969)}]{Bell}%
  \BibitemOpen
  \bibfield  {author} {\bibinfo {author} {\bibfnamefont {J.~S.}\ \bibnamefont
  {Bell}}\ and\ \bibinfo {author} {\bibfnamefont {R.}~\bibnamefont {Jackiw}},\
  }\href {\doibase 10.1007/BF02823296} {\bibfield  {journal} {\bibinfo
  {journal} {Il Nuovo Cimento A}\ }\textbf {\bibinfo {volume} {67}},\ \bibinfo
  {pages} {47} (\bibinfo {year} {1969})}\BibitemShut {NoStop}%
\bibitem [{\citenamefont {Nielsen}\ and\ \citenamefont
  {Ninomiya}(1981)}]{Nielsen}%
  \BibitemOpen
  \bibfield  {author} {\bibinfo {author} {\bibfnamefont {H.~B.}\ \bibnamefont
  {Nielsen}}\ and\ \bibinfo {author} {\bibfnamefont {M.}~\bibnamefont
  {Ninomiya}},\ }\href {\doibase 10.1016/0550-3213(81)90361-8} {\bibfield
  {journal} {\bibinfo  {journal} {Nucl. Phys. B}\ }\textbf {\bibinfo {volume}
  {185}},\ \bibinfo {pages} {20} (\bibinfo {year} {1981})}\BibitemShut
  {NoStop}%
\bibitem [{\citenamefont {Son}\ and\ \citenamefont {Spivak}(2013)}]{SonSpivak}%
  \BibitemOpen
  \bibfield  {author} {\bibinfo {author} {\bibfnamefont {D.~T.}\ \bibnamefont
  {Son}}\ and\ \bibinfo {author} {\bibfnamefont {B.~Z.}\ \bibnamefont
  {Spivak}},\ }\href {\doibase 10.1103/PhysRevB.88.104412} {\bibfield
  {journal} {\bibinfo  {journal} {Phys. Rev. B}\ }\textbf {\bibinfo {volume}
  {88}},\ \bibinfo {pages} {104412} (\bibinfo {year} {2013})}\BibitemShut
  {NoStop}%
\bibitem [{\citenamefont {Burkov}(2014)}]{PhysRevLett.113.247203}%
  \BibitemOpen
  \bibfield  {author} {\bibinfo {author} {\bibfnamefont {A.~A.}\ \bibnamefont
  {Burkov}},\ }\href {\doibase 10.1103/PhysRevLett.113.247203} {\bibfield
  {journal} {\bibinfo  {journal} {Phys. Rev. Lett.}\ }\textbf {\bibinfo
  {volume} {113}},\ \bibinfo {pages} {247203} (\bibinfo {year}
  {2014})}\BibitemShut {NoStop}%
\bibitem [{\citenamefont {Xiong}\ \emph {et~al.}(2015)\citenamefont {Xiong},
  \citenamefont {Kushwaha}, \citenamefont {Liang}, \citenamefont {Krizan},
  \citenamefont {Hirschberger}, \citenamefont {Wang}, \citenamefont {Cava},\
  and\ \citenamefont {Ong}}]{Xiong413}%
  \BibitemOpen
  \bibfield  {author} {\bibinfo {author} {\bibfnamefont {J.}~\bibnamefont
  {Xiong}}, \bibinfo {author} {\bibfnamefont {S.~K.}\ \bibnamefont {Kushwaha}},
  \bibinfo {author} {\bibfnamefont {T.}~\bibnamefont {Liang}}, \bibinfo
  {author} {\bibfnamefont {J.~W.}\ \bibnamefont {Krizan}}, \bibinfo {author}
  {\bibfnamefont {M.}~\bibnamefont {Hirschberger}}, \bibinfo {author}
  {\bibfnamefont {W.}~\bibnamefont {Wang}}, \bibinfo {author} {\bibfnamefont
  {R.~J.}\ \bibnamefont {Cava}}, \ and\ \bibinfo {author} {\bibfnamefont
  {N.~P.}\ \bibnamefont {Ong}},\ }\href {\doibase 10.1126/science.aac6089}
  {\bibfield  {journal} {\bibinfo  {journal} {Science}\ }\textbf {\bibinfo
  {volume} {350}},\ \bibinfo {pages} {413} (\bibinfo {year}
  {2015})}\BibitemShut {NoStop}%
\bibitem [{\citenamefont {Yang}\ \emph {et~al.}()\citenamefont {Yang},
  \citenamefont {Liu}, \citenamefont {Sun}, \citenamefont {Peng}, \citenamefont
  {Yang}, \citenamefont {Zhang}, \citenamefont {Zhou}, \citenamefont {Zhang},
  \citenamefont {Guo}, \citenamefont {Rahn}, \citenamefont {Prabhakaran},
  \citenamefont {Hussain}, \citenamefont {Mo}, \citenamefont {Felser},
  \citenamefont {Yan},\ and\ \citenamefont {Chen}}]{Felser}%
  \BibitemOpen
  \bibfield  {author} {\bibinfo {author} {\bibfnamefont {L.}~\bibnamefont
  {Yang}}, \bibinfo {author} {\bibfnamefont {Z.}~\bibnamefont {Liu}}, \bibinfo
  {author} {\bibfnamefont {Y.}~\bibnamefont {Sun}}, \bibinfo {author}
  {\bibfnamefont {H.}~\bibnamefont {Peng}}, \bibinfo {author} {\bibfnamefont
  {H.}~\bibnamefont {Yang}}, \bibinfo {author} {\bibfnamefont {T.}~\bibnamefont
  {Zhang}}, \bibinfo {author} {\bibfnamefont {B.}~\bibnamefont {Zhou}},
  \bibinfo {author} {\bibfnamefont {Y.}~\bibnamefont {Zhang}}, \bibinfo
  {author} {\bibfnamefont {Y.}~\bibnamefont {Guo}}, \bibinfo {author}
  {\bibfnamefont {M.}~\bibnamefont {Rahn}}, \bibinfo {author} {\bibfnamefont
  {D.}~\bibnamefont {Prabhakaran}}, \bibinfo {author} {\bibfnamefont
  {Z.}~\bibnamefont {Hussain}}, \bibinfo {author} {\bibfnamefont {S.-K.}\
  \bibnamefont {Mo}}, \bibinfo {author} {\bibfnamefont {C.}~\bibnamefont
  {Felser}}, \bibinfo {author} {\bibfnamefont {B.}~\bibnamefont {Yan}}, \ and\
  \bibinfo {author} {\bibfnamefont {Y.}~\bibnamefont {Chen}},\ }\href@noop {}
  {\bibinfo  {journal} {arXiv:1507.00521}\ }\BibitemShut {NoStop}%
\bibitem [{\citenamefont {Cortijo}()}]{Cort}%
  \BibitemOpen
\bibfield  {journal} {  }\bibfield  {author} {\bibinfo {author} {\bibfnamefont
  {A.}~\bibnamefont {Cortijo}},\ }\href@noop {} {\bibinfo  {journal}
  {arXiv:1610.06177}\ }\BibitemShut {NoStop}%
\bibitem [{\citenamefont {Morimoto}\ \emph {et~al.}()\citenamefont {Morimoto},
  \citenamefont {Zhong}, \citenamefont {Orenstein},\ and\ \citenamefont
  {Moore}}]{Moore1}%
  \BibitemOpen
\bibfield  {journal} {  }\bibfield  {author} {\bibinfo {author} {\bibfnamefont
  {T.}~\bibnamefont {Morimoto}}, \bibinfo {author} {\bibfnamefont
  {S.}~\bibnamefont {Zhong}}, \bibinfo {author} {\bibfnamefont
  {J.}~\bibnamefont {Orenstein}}, \ and\ \bibinfo {author} {\bibfnamefont
  {J.~E.}\ \bibnamefont {Moore}},\ }\href@noop {} {\bibinfo  {journal}
  {arXiv:1609.05932}\ }\BibitemShut {NoStop}%
\bibitem [{\citenamefont {Xu}\ \emph {et~al.}(2016)\citenamefont {Xu},
  \citenamefont {Dai}, \citenamefont {Zhao}, \citenamefont {Wang},
  \citenamefont {Yang}, \citenamefont {Zhang}, \citenamefont {Liu},
  \citenamefont {Xiao}, \citenamefont {Chen}, \citenamefont {Taylor},
  \citenamefont {Yarotski}, \citenamefont {Prasankumar},\ and\ \citenamefont
  {Qiu}}]{Weyl_optics1}%
  \BibitemOpen
\bibfield  {journal} {  }\bibfield  {author} {\bibinfo {author} {\bibfnamefont
  {B.}~\bibnamefont {Xu}}, \bibinfo {author} {\bibfnamefont {Y.~M.}\
  \bibnamefont {Dai}}, \bibinfo {author} {\bibfnamefont {L.~X.}\ \bibnamefont
  {Zhao}}, \bibinfo {author} {\bibfnamefont {K.}~\bibnamefont {Wang}}, \bibinfo
  {author} {\bibfnamefont {R.}~\bibnamefont {Yang}}, \bibinfo {author}
  {\bibfnamefont {W.}~\bibnamefont {Zhang}}, \bibinfo {author} {\bibfnamefont
  {J.~Y.}\ \bibnamefont {Liu}}, \bibinfo {author} {\bibfnamefont
  {H.}~\bibnamefont {Xiao}}, \bibinfo {author} {\bibfnamefont {G.~F.}\
  \bibnamefont {Chen}}, \bibinfo {author} {\bibfnamefont {A.~J.}\ \bibnamefont
  {Taylor}}, \bibinfo {author} {\bibfnamefont {D.~A.}\ \bibnamefont
  {Yarotski}}, \bibinfo {author} {\bibfnamefont {R.~P.}\ \bibnamefont
  {Prasankumar}}, \ and\ \bibinfo {author} {\bibfnamefont {X.~G.}\ \bibnamefont
  {Qiu}},\ }\href {\doibase 10.1103/PhysRevB.93.121110} {\bibfield  {journal}
  {\bibinfo  {journal} {Phys. Rev. B}\ }\textbf {\bibinfo {volume} {93}},\
  \bibinfo {pages} {121110} (\bibinfo {year} {2016})}\BibitemShut {NoStop}%
\bibitem [{\citenamefont {Sushkov}\ \emph {et~al.}(2015)\citenamefont
  {Sushkov}, \citenamefont {Hofmann}, \citenamefont {Jenkins}, \citenamefont
  {Ishikawa}, \citenamefont {Nakatsuji}, \citenamefont {Das~Sarma},\ and\
  \citenamefont {Drew}}]{Weyl_optics2}%
  \BibitemOpen
  \bibfield  {author} {\bibinfo {author} {\bibfnamefont {A.~B.}\ \bibnamefont
  {Sushkov}}, \bibinfo {author} {\bibfnamefont {J.~B.}\ \bibnamefont
  {Hofmann}}, \bibinfo {author} {\bibfnamefont {G.~S.}\ \bibnamefont
  {Jenkins}}, \bibinfo {author} {\bibfnamefont {J.}~\bibnamefont {Ishikawa}},
  \bibinfo {author} {\bibfnamefont {S.}~\bibnamefont {Nakatsuji}}, \bibinfo
  {author} {\bibfnamefont {S.}~\bibnamefont {Das~Sarma}}, \ and\ \bibinfo
  {author} {\bibfnamefont {H.~D.}\ \bibnamefont {Drew}},\ }\href {\doibase
  10.1103/PhysRevB.92.241108} {\bibfield  {journal} {\bibinfo  {journal} {Phys.
  Rev. B}\ }\textbf {\bibinfo {volume} {92}},\ \bibinfo {pages} {241108}
  (\bibinfo {year} {2015})}\BibitemShut {NoStop}%
\bibitem [{\citenamefont {Glazov}\ and\ \citenamefont
  {Ganichev}(2014)}]{Misha}%
  \BibitemOpen
  \bibfield  {author} {\bibinfo {author} {\bibfnamefont {M.}~\bibnamefont
  {Glazov}}\ and\ \bibinfo {author} {\bibfnamefont {S.}~\bibnamefont
  {Ganichev}},\ }\href
  {http://www.sciencedirect.com/science/article/pii/S0370157313003785}
  {\bibfield  {journal} {\bibinfo  {journal} {Physics Reports}\ }\textbf
  {\bibinfo {volume} {535}},\ \bibinfo {pages} {101} (\bibinfo {year}
  {2014})}\BibitemShut {NoStop}%
\bibitem [{\citenamefont {Son}\ and\ \citenamefont
  {Yamamoto}(2013)}]{PhysRevD.87.085016}%
  \BibitemOpen
  \bibfield  {author} {\bibinfo {author} {\bibfnamefont {D.~T.}\ \bibnamefont
  {Son}}\ and\ \bibinfo {author} {\bibfnamefont {N.}~\bibnamefont {Yamamoto}},\
  }\href {\doibase 10.1103/PhysRevD.87.085016} {\bibfield  {journal} {\bibinfo
  {journal} {Phys. Rev. D}\ }\textbf {\bibinfo {volume} {87}},\ \bibinfo
  {pages} {085016} (\bibinfo {year} {2013})}\BibitemShut {NoStop}%
\bibitem [{\citenamefont {Xiao}\ \emph {et~al.}(2010)\citenamefont {Xiao},
  \citenamefont {Chang},\ and\ \citenamefont {Niu}}]{RevModPhys.82.1959}%
  \BibitemOpen
  \bibfield  {author} {\bibinfo {author} {\bibfnamefont {D.}~\bibnamefont
  {Xiao}}, \bibinfo {author} {\bibfnamefont {M.-C.}\ \bibnamefont {Chang}}, \
  and\ \bibinfo {author} {\bibfnamefont {Q.}~\bibnamefont {Niu}},\ }\href
  {\doibase 10.1103/RevModPhys.82.1959} {\bibfield  {journal} {\bibinfo
  {journal} {Rev. Mod. Phys.}\ }\textbf {\bibinfo {volume} {82}},\ \bibinfo
  {pages} {1959} (\bibinfo {year} {2010})}\BibitemShut {NoStop}%
\bibitem [{\citenamefont {Zyuzin}()}]{Vova_Weyl}%
  \BibitemOpen
  \bibfield  {author} {\bibinfo {author} {\bibfnamefont {V.~A.}\ \bibnamefont
  {Zyuzin}},\ }\href@noop {} {\bibinfo  {journal} {arXiv:1608.01286}\
  }\BibitemShut {NoStop}%
\end{thebibliography}%
\appendix
\begin{widetext}
\section{Supplemental Material: Derivation of SHG}
\subsection{Main definitions and equilibrium solution}
Kinetic equation has the form
\begin{eqnarray}
\nonumber
\frac{\partial n_{\mathbf{k},s}}{\partial t} &+& D^{-1}_{\mathbf{k},s}\left[ e\tilde{\mathbf{E}}+ \frac{e}{c} \mathbf{v}_s\times \mathbf{B} + \frac{e^2}{c}( \tilde{\mathbf{E}}\cdot \mathbf{B})\mathbf{\Omega}_{\mathbf{k},s}\right]\cdot\frac{\partial n_{\mathbf{k},s}}{\partial \mathbf{k}} +D^{-1}_{\mathbf{k},s}\left[ \mathbf{v}_s + e\tilde{\mathbf{E}}\times\mathbf{\Omega}_{\mathbf{k},s} + \frac{e}{c}(\mathbf{v}_s\cdot\mathbf{\Omega}_{\mathbf{k}, s})\mathbf{B}\right]\cdot\frac{\partial n_{\mathbf{k},s}}{\partial \mathbf{r}}
\\
&=& \frac{\langle n_{\mathbf{k},s}\rangle - n_{\mathbf{k},s}}{\tau} - \frac{s}{\tau_v}\Lambda,
\end{eqnarray}
where 
\begin{equation}
D_{\mathbf{k},s}=1+\frac{e}{c}\mathbf{B}\cdot\mathbf{\Omega}_{\mathbf{k}, s},~~~e\tilde{\mathbf{E}} = e\mathbf{E}-\partial \mathcal{E}_{\mathbf{k},s}/\partial \mathbf{r},~~~\mathcal{E}_{\mathbf{k},s} = v k\left(1 - \frac{e}{c}\mathbf{B}\cdot\mathbf{\Omega}_{\mathbf{k},s}\right).
\end{equation}
Given that $\mathbf{\Omega}_{\mathbf{k},s}=s\hat{\mathbf{k}}/2k^2$ the velocity of the wavepacket is obtained as
\begin{equation}
\mathbf{v}_s = \frac{\partial}{\partial \mathbf{k}}\mathcal{E}_{\mathbf{k},s} = v\hat{\mathbf{k}}\left[1+\frac{2e}{c}(\mathbf{B}\cdot\mathbf{\Omega}_{\mathbf{k},s})\right] -  \frac{sve}{2ck}\mathbf{B}.
\end{equation}
The current density is given by
\begin{eqnarray}
&\mathbf{J}^{(2)}& = e\sum_{s=\pm} \int \frac{d^3k}{(2\pi)^3} \bigg\{\bigg[v\hat{\mathbf{k}}\frac{2e}{c}(\mathbf{B}_0\cdot\mathbf{\Omega}_{\mathbf{k},s}) +e\mathbf{E}_0\times\mathbf{\Omega}_{\mathbf{k},s}
+  \frac{sve^2}{2c^2k^2}[(\mathbf{B}_c\cdot\mathbf{\Omega}_{\mathbf{k},s})\mathbf{B}_0+(\mathbf{B}_0\cdot\mathbf{\Omega}_{\mathbf{k},s})\mathbf{B}_c]
\\\nonumber
&+& ivk\frac{e}{c}(\mathbf{B}_0\cdot\mathbf{\Omega}_{\mathbf{k},s})[\mathbf{q}\times \mathbf{\Omega}_{\mathbf{k},s}]\bigg]n^{(1)}_{\mathbf{k},s}-2 i vk [\mathbf{q} \times\mathbf{\Omega}_{\mathbf{k},s} ]n^{(2)}_{\mathbf{k},s}+  v\hat{\mathbf{k}}(1+\frac{2e}{c}\mathbf{B}_c\cdot\mathbf{\Omega}_{\mathbf{k},s})  n^{(2)}_{\mathbf{k},s} \bigg\}.
\end{eqnarray}
We can write expression for the current density in the form
\begin{equation}\label{Cur_split_appendix}
\mathbf{J}^{(2)} = \mathbf{J}^{(2)}_1+\mathbf{J}^{(2)}_2,
\end{equation} 
such that $\mathbf{J}^{(2)}_{1,2}$ is a function of $n^{(1,2)}_{\mathbf{k},s}$, respectively. 
We search for the solution of kinetic equation perturbatively in powers of the amplitude of the incident electromagnetic field. 
We expand the distribution function around the equilibrium Fermi-Dirac distribution
\begin{equation}
n^{(0)}_{\mathbf{k},s} = [1+e^{\left(v k(1 - \frac{e}{c}\mathbf{B}_c\cdot\mathbf{\Omega}_{\mathbf{k},s}) -\mu\right)/T}]^{-1}
\end{equation}
as follows
\begin{eqnarray} \label{expansion}
n_{\mathbf{k},s}=n^{(0)}_{\mathbf{k},s} + [n^{(1)}_{\mathbf{k},s} e^{i(\mathbf{q}\cdot\mathbf{r}-\omega t)}+n^{(2)}_{\mathbf{k},s}e^{2i(\mathbf{q}\cdot\mathbf{r}-\omega t)} + \textrm{c.c.}],
\end{eqnarray}
where $n^{(j)}_{\mathbf{k},s}$ is proportional to $j$-power of the amplitude of electromagnetic field.
We similarly expand the function 
\begin{equation}
\Lambda=\Lambda^{(0)} +[\Lambda^{(1)} e^{i(\mathbf{q}\cdot\mathbf{r}-\omega t)}+\Lambda^{(2)}e^{2i(\mathbf{q}\cdot\mathbf{r}-\omega t)} + \textrm{c.c.}]
\end{equation}
and observe that $\Lambda^{(0)}=0$.

\subsection{First order correction}
Let us introduce new variables $(k_{\perp}\cos(\phi), k_{\perp}\sin(\phi), k_z)=(k_x,k_y,k_z) $, which allows to rewrite 
\begin{equation}
\frac{ev}{c}[\hat{\mathbf{k}}\times\mathbf{B}_c]\cdot \frac{\partial n}{\partial \mathbf{k}} = \tilde{\omega}_c \frac{\partial n}{\partial \phi}
\end{equation}
and $D_{\mathbf{k},s} = 1-s\tilde{k}_z \tilde{\omega}_c/2vk$, where $\tilde{\omega}_c=-ev B_c/ck$ is the cyclotron frequency of electron with momentum k. Note that in the main text we use $\omega_c = \frac{vk}{\mu}\tilde{\omega}_c$. Working in the long wave limit $\omega\gg vq$ we will consider wave vector dependent contributions to the distribution function as a perturbation. We search for a solution in the limit when the distribution function is spacial independent function. 
Thus, kinetic equation for the first order correction $n^{(1)}_{\mathbf{k},s}$ has the form
\begin{equation}\label{first correction}
\bigg[-i\omega n^{(1)}_{\mathbf{k},s} + \frac{s\Lambda^{(1)}}{\tau_v} +\frac{n^{(1)}_{\mathbf{k},s}-\langle n^{(1)}_{\mathbf{k},s}\rangle}{\tau}\bigg]D_{\mathbf{k},s}
+\omega_c\tilde{D}_{\mathbf{k},s}\frac{\partial  n^{(1)}_{\mathbf{k},s}}{\partial\phi}
=-\bigg[e\mathbf{E}_{0}+\frac{e^2}{c}(\mathbf{E}_0\cdot\mathbf{B}_c)\mathbf{\Omega}_{\mathbf{k},s} +\frac{ev}{c} [\hat{\mathbf{k}}\times \mathbf{B}_0]\bigg]\cdot\frac{\partial  n^{(0)}_{\mathbf{k},s}}{\partial \mathbf{k}},
\end{equation}
where we introduce $\tilde{D}_{\mathbf{k},s}=2{D}_{\mathbf{k},s}-1$ for brevity. Right hand side of this equation contains contributions from magnetic field $\mathbf{B}_c$, among which first term originates from the Berry phase contribution while second term is due to orbital magnetic moment. Noting that $n^{(0)}_{\mathbf{k},s}$ depends on $\mathbf{B}_c\cdot \hat{k} \equiv B_c \hat{k}_z$, we can rewrite Eq. \ref{first correction} in the form
\begin{equation}\label{N1_appendix}
\bigg[-i\omega n^{(1)}_{\mathbf{k},s} + \frac{s\Lambda^{(1)}}{\tau_v} +\frac{n^{(1)}_{\mathbf{k},s}-\langle n^{(1)}_{\mathbf{k},s}\rangle}{\tau}\bigg]D_{\mathbf{k},s}
+\tilde{\omega}_c\tilde{D}_{\mathbf{k},s}\frac{\partial  n^{(1)}_{\mathbf{k},s}}{\partial\phi}
=-e\bigg[\mathbf{E}_0\cdot\hat{\mathbf{k}} +\frac{se}{2ck^2} \left(\mathbf{E}_0+\frac{v}{c}\mathbf{B}_0\times\hat{\mathbf{k}}_{\perp}\right)\cdot\mathbf{B}_c\bigg]\frac{\partial  
n^{(0)}_{\mathbf{k},s}}{\partial k},
\end{equation}
where $\mathbf{E}_0\cdot\hat{\mathbf{k}}  = E_{0,z}\hat{k}_z + E_{0,x}\hat{k}_{\perp}\cos{(\phi)}+E_{0,y}\hat{k}_{\perp}\sin{(\phi)} $.
The separation of variables along and transverse to the direction of magnetic field $\mathbf{B}_c$ allows to solve this equation exactly.
It is convenient to write the first order correction to the distribution function as
$n^{(1)}_{\mathbf{k},s}=n^{(1,\|)}_{\mathbf{k},s}+n^{(1,\bot)}_{\mathbf{k},s}$, where
\begin{subequations}\label{n1_Solution}
\begin{align}
n^{(1,\|)}_{\mathbf{k},s} &= \frac{1/2-\tau/\tau_v}{1-i\omega\tau}s\Lambda^{(1)}- \frac{ e\tau E_{0,z} }{(1-i\omega\tau)D_{\mathbf{k},s}}\left[\hat{k}_z-\frac{s\tilde{\omega}_c}{2vk}\right]\frac{\partial n^{(0)}_{\mathbf{k},s} }{\partial k},\\
n^{(1,\bot)}_{\mathbf{k},s}&=-e\tau
\left[\frac{\tilde{\omega}_c\tau\tilde{D}_{\mathbf{k},s}([\mathbf{E}_0\times \hat{\mathbf{k}}]\cdot\hat{z}+\frac{s \tilde{\omega}_c}{2ck}\mathbf{B}_{0}\cdot\hat{\mathbf{k}}_{\perp})}{(\tilde{\omega}_c\tau)^2\tilde{D}_{\mathbf{k},s}^2+(1-i\omega\tau)^2D_{\mathbf{k},s}^2}
+\frac{(1-i\omega\tau)D_{\mathbf{k},s}(\mathbf{E}_0\cdot\hat{\mathbf{k}}_{\perp}+  \frac{s \tilde{\omega}_c}{2ck}[\hat{\mathbf{k}}\times\mathbf{B}_{0}]\cdot\hat{z}  )}{(\tilde{\omega}_c\tau)^2\tilde{D}_{\mathbf{k},s}^2+(1-i\omega\tau)^2D_{\mathbf{k},s}^2}\right]\frac{\partial n^{(0)}_{\mathbf{k},s} }{\partial k}.
\end{align}
\end{subequations}
Here we separate contributions to $n^{(1)}_{\mathbf{k},s}$ from parallel and perpendicular components of the electric field $\mathbf{E}_0$ with respect to $\mathbf{B}_c$.
By integrating Eq. \ref{N1_appendix} over directions of momentum $\mathbf{k}$, we find the first order correction to the inter-valley coupling in the form
\begin{eqnarray}
\Lambda^{(1)}=\frac{-e\tau_v/2}{1-i\omega\tau_v/2}\int\frac{d\vec{\Theta}}{4\pi}\sum_{s=\pm}\bigg[\frac{e}{2c k}(\mathbf{E}_{0}\cdot \mathbf{B}_c)
+sE_{0,z}\hat{k}_z\bigg]\frac{\partial n^{(0)}_{\mathbf{k},s}}{\partial k}.
\end{eqnarray}
Note that both orbital magnetic moment $\mathbf{m}_{\mathbf{k},s}$ term through the magnetic field dependence of the distribution function $n^{(0)}_{\mathbf{k},s}$ and Berry phase related term $\mathbf{E}_{0}\cdot \mathbf{B}_c$ contribute to $\Lambda^{(1)}$.

\subsection{SHG from first order correction}
Contribution to SHG from $n^{(1)}_{\mathbf{k},s}$ is defined as
\begin{eqnarray}\label{n1_current}
\mathbf{J}^{(2)}_1 &=& e\sum_{s=\pm} \int \frac{d^3k}{(2\pi)^3} \bigg\{v\hat{\mathbf{k}}\frac{2e}{c}(\mathbf{B}_0\cdot\mathbf{\Omega}_{\mathbf{k},s}) +e\mathbf{E}_0\times\mathbf{\Omega}_{\mathbf{k},s}
+  \frac{sve^2}{2c^2k^2}\bigg[(\mathbf{B}_c\cdot\mathbf{\Omega}_{\mathbf{k},s})\mathbf{B}_0+(\mathbf{B}_0\cdot\mathbf{\Omega}_{\mathbf{k},s})\mathbf{B}_c\bigg] \bigg\} n^{(1)}_{\mathbf{k},s}.
\end{eqnarray}
We observe here that all terms are determined by the topological electronic band structure.
Substituting solution for $n^{(1)}_{\mathbf{k},s}$ given by Eq.~\ref{n1_Solution} into Eq.~\ref{n1_current} we find
\begin{eqnarray}\nonumber\label{Current_n1}
\mathbf{J}^{(2)}_1 &=& - \frac{ve^4 I}{3c^2} \left[\tau_v\frac{(\mathbf{E}_0\cdot\mathbf{B}_c)\mathbf{B}_0}{1-i\omega \tau_v/2}+\frac{\tau}{2}\frac{(\omega_c\tau)^2 E_{0,z}(B_c\mathbf{B}_0+B_{0,z}\mathbf{B}_c)}{(1-i\omega\tau)((1-i\omega\tau)^2+(\omega_c\tau)^2)}\right]
+\frac{v e^4 I }{15c^2}\frac{\tau}{(1-i\omega\tau)}\frac{B_c}{(1-i\omega\tau)^2+(\omega_c\tau)^2}\\
&\times&\bigg[(\omega_c\tau)^2E_{0,z}(\mathbf{B}_0+2\mathbf{B}_{0,z})+(1-i\omega\tau)^2(E_{0,z}\mathbf{B}_{0,\perp}+B_{0,z}\mathbf{E}_{0,\perp})+\omega_c\tau(1-i\omega\tau)[\mathbf{E}_{0,\perp}\times \mathbf{B}_{0,\perp}]\bigg],
\end{eqnarray}
where
\begin{equation}
I = \int_{k_0}^{\infty} \frac{dk}{2\pi^2 k^2}\frac{\partial n^{(0)}_{\mathbf{k},s}}{\partial k}\bigg|_{B_c=0} \approx -\frac{v}{2\pi^2\mu^2}\left[1+\frac{\mu^2}{vk_0 T}e^{-\mu/T}\right],
\end{equation}
in which $vk_0$ is the low energy cut-off and $\mu\gg T, vk_0$ is assumed. Here one might distinguish between contributions to SHG that originate solely from the intra-valley and inter-valley scattering processes. We also note that $\mathbf{J}^{(2)}_1=0$ if either $B_0=0$ or $B_c=0$.
Assuming the limit $B_c\rightarrow 0$ we obtain
\begin{eqnarray}
\mathbf{J}^{(2)}_1 = - \frac{2ve^4 I}{3c^2} \frac{\tau_v/2}{1-i\omega \tau_v/2}(\mathbf{E}_0\cdot\mathbf{B}_c)\mathbf{B}_0 + \frac{ve^4 I}{15 c^2}\frac{\tau}{1-i\omega\tau}(E_{0,z}\mathbf{B}_{0,\perp}+B_{0,z}\mathbf{E}_{0,\perp})B_c.
\end{eqnarray}
First term describes the effect of chiral anomaly, second originates from orbital magnetic moment.

\subsection{Second order correction}
Assuming condition for transverse electromagnetic wave, $\mathbf{E}_0\cdot\mathbf{B}_0=0$, we write down equation for the second order correction $n_{\mathbf{k},s}^{(2)}$ in the form
\begin{eqnarray}\nonumber
&&D_{\mathbf{k},s}\left[\left(-2i\omega +\frac{1}{\tau}\right)n_{\mathbf{k},s}^{(2)}+\frac{s\Lambda^{(2)}}{\tau_{v}}-\frac{\langle n_{\mathbf{k},s}^{(2)}\rangle}{\tau}\right]+\tilde{D}_{\mathbf{k},s}\frac{ev}{c}[\hat{\mathbf{k}}\times\mathbf{B}_c]\cdot\frac{\partial n_{\mathbf{k},s}^{(2)}}{\partial \mathbf{k}}\\\nonumber
&=&
\frac{e^2 \mathbf{B}_0\cdot\mathbf{\Omega}_{\mathbf{k},s}}{cD_{\mathbf{k},s}}\bigg[\mathbf{E}_0+ \frac{e}{c}(\mathbf{E}_0\cdot\mathbf{B}_c)\mathbf{\Omega}_{\mathbf{k},s}\bigg]\cdot\frac{\partial n^{(0)}_{\mathbf{k},s}}{\partial \mathbf{k}}-\frac{ev}{c}[\hat{\mathbf{k}}\times\mathbf{B}_0]\cdot \frac{\partial n_{\mathbf{k},s}^{(1)}}{\partial \mathbf{k}}\\
&+&\frac{ve^2 \mathbf{B}_0\cdot\mathbf{\Omega}_{\mathbf{k},s}}{c^2D_{\mathbf{k},s}} [\hat{\mathbf{k}}\times \mathbf{B}_{0}]\cdot\frac{\partial n^{(0)}_{\mathbf{k},s}}{\partial \mathbf{k}} -\left[ e\mathbf{E}_0+\frac{e^2v}{c^2}(\mathbf{\Omega}_{\mathbf{k},s}\cdot\mathbf{B}_0)[\hat{\mathbf{k}}\times\mathbf{B}_c] +\frac{e^2}{c}(\mathbf{E}_0\cdot\mathbf{B}_c)\mathbf{\Omega}_{\mathbf{k},s} \right]\cdot \frac{\partial n_{\mathbf{k},s}^{(1)}}{\partial \mathbf{k}}.
\end{eqnarray}
We observe that the part of the solution that is determined by the terms on the third line of this equation does not contribute to SHG within our linear in $\frac{e}{c}\mathbf{B}_c\cdot\mathbf{\Omega}_{\mathbf{k},s}$ approximation. Thus, we ignore them from our consideration for brevity. We emphasize that for $B_0=0$ the right hand side of this equation is given by $- e\mathbf{E}_0\cdot\partial n_{\mathbf{k},s}^{(1)} /\partial \mathbf{k}$. Using the parity of the integrand with respect to $s\mathbf{k}$ it can be shown that SHG vanishes in this case.

We then find that $\Lambda^{(2)}=0$ together with $\langle n_{\mathbf{k},s}^{(2)}\rangle =0$. Noting that $\tilde{D}_{\mathbf{k},s}/D_{\mathbf{k},s}\approx D_{\mathbf{k},s}$ we rewrite equation for the second order correction in the linear order in magnetic filed $B_c$ in the form
\begin{equation}
\left[-2i\omega+\frac{1}{\tau}\right]\bar{n}_{\mathbf{k},s}^{(2)}+\tilde{\omega}_cD_{\mathbf{k},s}\frac{\partial \bar{n}_{\mathbf{k},s}^{(2)}}{\partial \phi} =
\frac{e^2 \mathbf{B}_0\cdot\mathbf{\Omega}_{\mathbf{k},s}}{cD^2_{\mathbf{k},s}}\bigg[\mathbf{E}_0+ \frac{e}{c}(\mathbf{E}_0\cdot\mathbf{B}_c)\mathbf{\Omega}_{\mathbf{k},s}\bigg]\cdot\frac{\partial n^{(0)}_{\mathbf{k},s}}{\partial \mathbf{k}}-\frac{ev}{c}[\hat{\mathbf{k}}\times\mathbf{B}_0]\cdot \frac{\partial n_{\mathbf{k},s}^{(1)}}{\partial \mathbf{k}}.
\end{equation}
We observe that the first term on the rhs of this equation originates from the topological electronic band structure, which is a correction to the $EB$ mechanism of SHG described by the last term.
Quite lengthly solution of this equation is given by
\begin{eqnarray}\nonumber\label{n2_appendix}
\bar{n}_{\mathbf{k},s}^{(2)} &=& \frac{\tau}{1-2i\omega\tau}\frac{e^3(\mathbf{E}_0 \cdot\mathbf{B}_c)}{(2ck^2)^2} \frac{\partial n^{(0)}_{\mathbf{k},s}/\partial k}{(\tilde{\omega}_c\tau)^2 +(1-2i\omega\tau)^2}\left[(1-2i\omega\tau)^2 \mathbf{B}_0\cdot\hat{\mathbf{k}} +(\tilde{\omega}_c\tau)^2 B_{0,z}\hat{k}_z + \tilde{\omega}_c\tau(1-2i\omega\tau)[\mathbf{B}_0\times \hat{\mathbf{k}}]\cdot\hat{z} \right]\\\nonumber
&+& \frac{\tau}{1-2i\omega\tau}\frac{se^2}{2ck^2D_{\mathbf{k},s}^2}\frac{\partial n^{(0)}_{\mathbf{k},s}/\partial k}{(\tilde{\omega}_c\tau D_{\mathbf{k},s})^2 +(1-2i\omega\tau)^2}\bigg[ 
\tilde{\omega}_c\tau(1-2i\omega\tau)\hat{k}_z\bigg\{ B_{0,z}[\mathbf{E}_0\times\hat{\mathbf{k}}]\cdot\hat{z}+E_{0,z}[\mathbf{B}_0\times\hat{\mathbf{k}}]\cdot\hat{z} \bigg\}
\\\nonumber
&+&
(1-2i\omega\tau)^2\bigg\{\hat{k}_z\bigg(B_{0,z}(\mathbf{E}_0\cdot\hat{\mathbf{k}})+E_{0,z}(\mathbf{B}_0\cdot\hat{\mathbf{k}})-B_{0,z}E_{0,z}\hat{k}_z \bigg) +(\mathbf{B}_{0,\perp}\cdot\mathbf{E}_{0,\perp})\frac{\hat{k}^2_{\perp}}{2}\bigg\}
\\\nonumber
&+&(\tilde{\omega}_c\tau D_{\mathbf{k},s})^2\bigg\{ B_{0,z} E_{0,z}\hat{k}_z^2+(\mathbf{B}_{0,\perp}\cdot\mathbf{E}_{0,\perp})\frac{\hat{k}_{\perp}^2}{2}\bigg\}\bigg]
+F(\cos(2\phi),\sin(2\phi))
\\\nonumber
&+&\frac{e^2\tau^2v}{ck(1-i\omega\tau)(1-2i\omega\tau)}\frac{\partial n^{(0)}_{\mathbf{k},s}/\partial k}{(\tilde{\omega}_c\tau)^2+(1-i\omega\tau)^2}\frac{1}{(\tilde{\omega}_c\tau)^2+(1-2i\omega\tau)^2}
\\\nonumber
&\times&\bigg\{\hat{k}_z((\tilde{\omega}_c\tau)^2+(1-2i\omega\tau)^2)\left[\tilde{\omega}_c\tau(1-i\omega\tau)\mathbf{E}_{0,\perp}\cdot\mathbf{B}_{0,\perp}+(1-i\omega\tau)^2[\mathbf{B}_0\times\mathbf{E}_0]\cdot\hat{z}
\right]\\
&+&(1-2i\omega\tau)[\tilde{\omega}_c\tau[\mathbf{G}\times\hat{\mathbf{k}}]\cdot\hat{z} +(1-2i\omega\tau)(\mathbf{G}\cdot\hat{\mathbf{k}})\cdot\hat{z} ]
\bigg\},
\end{eqnarray}
where $F(\cos(2\phi),\sin(2\phi))$ is some function of $\cos(2\phi), \sin(2\phi)$, which will not contribute to SHG. We also introduce vector $\mathbf{G}=(G_x,G_y,0)$ for brevity, where 
\begin{subequations}
\begin{align}
&G_x=-\tilde{\omega}_c\tau(1-i\omega\tau)E_{0,x}B_{0,z}+(\tilde{\omega}_c\tau)^2E_{0,z}B_{0,y}+(1-i\omega\tau)^2[\mathbf{B}_0\times\mathbf{E}_0]\cdot\hat{x},\\
&G_x=-\tilde{\omega}_c\tau(1-i\omega\tau)E_{0,y}B_{0,z}-(\tilde{\omega}_c\tau)^2E_{0,z}B_{0,x}+(1-i\omega\tau)^2[\mathbf{B}_0\times\mathbf{E}_0]\cdot\hat{y}.
\end{align}
\end{subequations}

\subsection{SHG from second order correction}
Expression for SHG coming from from the second order correction is given by
\begin{eqnarray}
\mathbf{J}^{(2)}_2 = e\sum_{s=\pm} \int \frac{d^3k}{(2\pi)^3} v\hat{\mathbf{k}}\left(1+\frac{2e}{c}\mathbf{B}_c\cdot\mathbf{\Omega}_{\mathbf{k},s}\right)  n^{(2)}_{\mathbf{k},s}.
\end{eqnarray}
Substituting here $n^{(2)}_{\mathbf{k},s} \rightarrow\bar{n}^{(2)}_{\mathbf{k},s}$ we obtain
\begin{eqnarray}\label{Cur_n2_appendix}
\mathbf{J}^{(2)}_2 &=& \frac{ve^4 I}{6c^2} \frac{\tau(\mathbf{E}_0 \cdot\mathbf{B}_c)}{1-2i\omega\tau}\frac{(1-2i\omega\tau)^2 \mathbf{B}_0 +(\omega_c\tau)^2 B_{0,z}\hat{z} + \omega_c\tau(1-2i\omega\tau)[\hat{z}\times\mathbf{B}_0] }{(\omega_c\tau)^2 +(1-2i\omega\tau)^2}\\\nonumber
&+&
\frac{e^3v\tau}{2c(1-2i\omega\tau)} \sum_{s=\pm}\int \frac{d^3k}{(2\pi)^3} \frac{s\hat{\mathbf{k}} }{k^2}\frac{\partial n^{(0)}_{\mathbf{k},s}/\partial k}{(\tilde{\omega}_c\tau D_{\mathbf{k},s})^2 +(1-2i\omega\tau)^2}\bigg[ 
\tilde{\omega}_c\tau(1-2i\omega\tau)\hat{k}_z\bigg\{ B_{0,z}[\mathbf{E}_0\times\hat{\mathbf{k}}]\cdot\hat{z}+E_{0,z}[\mathbf{B}_0\times\hat{\mathbf{k}}]\cdot\hat{z} \bigg\}\\\nonumber
&+&
(1-2i\omega\tau)^2\bigg\{\hat{k}_z\bigg(B_{0,z}(\mathbf{E}_0\cdot\hat{\mathbf{k}})+E_{0,z}(\mathbf{B}_0\cdot\hat{\mathbf{k}})-B_{0,z}E_{0,z}\hat{k}_z \bigg) +(\mathbf{B}_{0,\perp}\cdot\mathbf{E}_{0,\perp})\hat{k}^2_{\perp}/2\bigg\}
\\\nonumber
&+&(\tilde{\omega}_c\tau D_{\mathbf{k},s})^2\bigg\{ B_{0,z} E_{0,z}\hat{k}_z^2+(\mathbf{B}_{0,\perp}\cdot\mathbf{E}_{0,\perp})\hat{k}_{\perp}^2/2\bigg\}\bigg]
\\\nonumber
&+&\frac{2e^3v^2\tau^2}{c(1-i\omega\tau)(1-2i\omega\tau)}
\int \frac{d^3k}{(2\pi)^3}
[(\tilde{\omega}_c\tau)^2+(1-i\omega\tau)^2]^{-1}[(\tilde{\omega}_c\tau)^2+(1-2i\omega\tau)^2]^{-1}
\\\nonumber
&\times&
\frac{\hat{\mathbf{k}}}{k}\bigg[(1-2i\omega\tau)^2\bigg\{ (1-i\omega\tau)^2\hat{\mathbf{k}}\cdot[\mathbf{B}_0\times\mathbf{E}_0]+(\tilde{\omega}_c\tau)^2E_{0,z}[\hat{\mathbf{k}}\times\mathbf{B}_0]\cdot\hat{z}
\\\nonumber
&-&\tilde{\omega}_c\tau(1-i\omega\tau)B_{0,z}\hat{\mathbf{k}}\cdot\mathbf{E}_0\bigg\}+(\tilde{\omega}_c\tau)^2\bigg\{-\tilde{\omega}_c\tau(1-i\omega\tau)E_{0,z}B_{0,z}\hat{k}_z +(1-i\omega\tau)^2\hat{k}_z[\mathbf{B}_0\times\mathbf{E}_0]\cdot\hat{z}\bigg\}\\\nonumber
&+&\tilde{\omega}_c\tau(1-2i\omega\tau)\bigg\{ -\tilde{\omega}_c\tau(1-i\omega\tau)B_{0,z}[\mathbf{E}_0\times\hat{\mathbf{k}}]\cdot\hat{z}+(\tilde{\omega}_c\tau)^2E_{0,z}(\hat{\mathbf{k}}_{\perp}\cdot\mathbf{B}_{0,\perp})
+(1-i\omega\tau)^2[[\mathbf{B}_0\times\mathbf{E}_0]\times\hat{\mathbf{k}}]\cdot\hat{z}
\bigg\}
\bigg]\frac{\partial n^{(0)}_{\mathbf{k},s}}{\partial k}\bigg|_{B_c=0}.
\end{eqnarray}
Solution of the second integral in the above expression describes $EB$-mechanism to SHG. Indeed, setting $B_c\rightarrow0$ one observes that the contribution to SHG is given by
\begin{eqnarray}\nonumber
\mathbf{J}^{(2)}_2 &=& \left(-\frac{ev^2B_0}{c\mu}\right) \frac{\sigma \tau}{(1-i\omega\tau)(1-2i\omega\tau)}\left\{[\hat{\mathbf{B}}_0\times\mathbf{E}_0] -\frac{\omega_c\tau}{(1-i\omega\tau)(1-2i\omega\tau)} \left[\frac{\hat{B}_{0,z}\mathbf{E}_0}{1-2i\omega\tau}+\frac{[\hat{\mathbf{B}}_c\times[\hat{\mathbf{B}}_{0}\times\mathbf{E}_0]]}{1-i\omega\tau}\right]\right\}\\
&+&\frac{ve^4 I}{6c^2} \frac{\tau}{1-2i\omega\tau} (\mathbf{E}_0 \cdot\mathbf{B}_c) \mathbf{B}_0.
\end{eqnarray}
where terms on the first line describe $EB$-mechanism to SHG, while term on the next line is a correction to this mechanism due to nontrivial topological electronic band structure. 

\subsection{SHG in the limit $B_c\rightarrow 0$}
Combining $\mathbf{J}^{(2)}_{1,2}$ in the limit $B_c\rightarrow 0$ we obtain
\begin{eqnarray}\nonumber
\mathbf{J}^{(2)} &=& \left(-\frac{ev^2B_0}{c\mu}\right) \frac{\sigma \tau}{(1-i\omega\tau)(1-2i\omega\tau)}\left\{[\hat{\mathbf{B}}_0\times\mathbf{E}_0] -\frac{\omega_c\tau}{(1-i\omega\tau)(1-2i\omega\tau)} \left[\frac{\hat{B}_{0,z}\mathbf{E}_0}{1-2i\omega\tau}+\frac{[\hat{\mathbf{B}}_c\times[\hat{\mathbf{B}}_{0}\times\mathbf{E}_0]]}{1-i\omega\tau}\right]\right\}\\
&-&\frac{2ve^4 I}{3c^2} \frac{\tau_v/2}{1-i\omega \tau_v/2}(\mathbf{E}_0\cdot\mathbf{B}_c)\mathbf{B}_0 + \frac{ve^4 I}{15 c^2}\frac{\tau}{1-i\omega\tau}(E_{0,z}\mathbf{B}_{0,\perp}+B_{0,z}\mathbf{E}_{0,\perp})B_c+\frac{ve^4 I}{6c^2} \frac{\tau}{1-2i\omega\tau} (\mathbf{E}_0 \cdot\mathbf{B}_c) \mathbf{B}_0.
\end{eqnarray}
We observe that in this limit topological band structure contribution to SHG, which originates from inter-valley scattering (first term on the second line), dominates over the intra-valley scattering contributions (other terms on the second line) if $\tau_v\gg \tau$. Inter-valley scattering contribution dominates over standard EB-mechanism if 
\begin{equation}
\frac{\tau_v}{\tau}\frac{\omega_c\tau}{(\mu\tau)^2}>1.
\end{equation}

\end{widetext}
\end{document}